\documentclass[aps, prl, twocolumn,showpacs]{revtex4}
\usepackage{epsfig}
\usepackage{graphicx}
\usepackage{amsmath}
\usepackage{amssymb}
\usepackage{amsbsy}
\usepackage{amsfonts}
\usepackage{dsfont}
\usepackage{color}

\newcommand{\beq}{\begin{eqnarray}}
\newcommand{\eeq}{\end{eqnarray}}
\newcommand\bsy{\mathbf}
\newcommand\trace{\mathop\mathrm{Tr}\nolimits}

\newcommand\Jxy{J_{xy}}
\newcommand\Jz{J_z}
\newcommand\re{\mathop\mathrm{Re}}
\newcommand\im{\mathop\mathrm{Im}}
\renewcommand\b{b}

\def\rank{\mathop\mathrm{rank}}
\def\kb{k_\mathrm{B}}

\begin{document}
\title{Highly Entangled Ground States in Tripartite Qubit Systems}
\author{Beat R\"othlisberger, J\"org Lehmann, D. S. Saraga, Philipp Traber, and Daniel Loss}
\affiliation{Department of Physics, University of
Basel, Klingelbergstrasse 82, CH-4056 Basel, Switzerland}
\date{\today}
\begin{abstract}
  We investigate the creation of highly entangled ground states in a
  system of three exchange-coupled qubits arranged in a
  ring geometry. Suitable magnetic field configurations
  yielding approximate GHZ and exact W ground states are identified. The entanglement in the system is studied at finite temperature in terms of the mixed-state tangle $\tau$. By generalizing a conjugate gradient optimization algorithm originally developed to evaluate the entanglement of formation, we demonstrate that $\tau$ can be calculated efficiently and with high precision. We identify the parameter regime for which the equilibrium entanglement of the tripartite system reaches its maximum.
\end{abstract}
\pacs{03.67.Mn, 03.65.Ud}

\maketitle

Entangled quantum systems have been the focus of numerous theoretical and
experimental investigations \cite{Coll01, Greenberger1989, Dur2000}. In
particular, entanglement has been identified as the primary resource
for quantum computation and communication~\cite{Nielsen2000a}.  
Compared to the case of a bipartite system, multipartite entanglement exhibits various new features. Notably, there are two
different equivalence classes of genuine three-qubit entanglement
\cite{Dur2000}, the representatives being any one of the two maximally entangled Greenberger-Horne-Zeilinger (GHZ)
states~\cite{Greenberger1989} $|\mathrm{GHZ}^{\pm}\rangle =
(|{\uparrow\uparrow\uparrow}\rangle \pm
|{\downarrow\downarrow\downarrow}\rangle)/\sqrt{2}$ on the one hand,
and the W state \cite{Dur2000} $|\mathrm{W}\rangle =
(|{\uparrow\uparrow\downarrow}\rangle +
|{\uparrow\downarrow\uparrow}\rangle +
|{\downarrow\uparrow\uparrow}\rangle)/\sqrt{3}$ on the other. 
The ability to realize both representatives in real physical systems is thus of high importance in the study of genuine tripartite entanglement. Particularly interesting is the GHZ state, as it represents the strongest quantum correlations possible in a system of three qubits. Furthermore, it is equivalent to the three-qubit cluster state used in one-way quantum computation \cite{Raussendorf2003}. It is favorable to obtain the GHZ and W states as the eigenstate of a suitable system, rather than by engineering them using quantum gates. In this Letter, we demonstrate the possibility of obtaining approximate GHZ and
exact W states as the ground state (g.s.) of three spin-qubits in a ring
geometry coupled via an anisotropic Heisenberg interaction. The use of quantum gates is therefore not required. Rather, the desired
states are achieved merely by cooling down to sufficiently low temperatures. We state all our results in terms of the exchange coupling strengths in order to keep our proposal open to a broad set of possible implementations of the qubits. We remark that, while Heisenberg models have been studied frequently in the context of entanglement \cite{Bose2003} (also with respect to entangled eigenstates \cite{Rajagopal2002}), this is the first time that highly entangled states are reported as the non-degenerate g.s.~of three exchange-coupled qubits. Our study inevitably involves the issue of quantifying entanglement \cite{Plenio2007, Mintert2005, Bennett1996, Coffman2000, Wei2003}: At finite temperatures, the mixing of the g.s.~with
excited states forces us to evaluate a mixed-state entanglement
measure (EM) in order to study the entanglement in the system meaningfully.
Computationally, this is a rather formidable task. We generalize a numerical
scheme that has originally been developed to compute the entanglement of formation (EOF) \cite{Bennett1996, Audenaert2001}. Our scheme can be used to evaluate \emph{any} mixed-state EM defined as a so-called convex roof \cite{Uhlmann2000}.

\textit{Model}.--- We assume that three spins $\bsy{S}_i$, with $S = 1/2$, are
located at the corners $i=1,2,3$ of an equilateral triangle lying in
the $xy$-plane. Their interaction is described by the anisotropic
Heisenberg Hamiltonian
\begin{equation}\label{anisotropic hamiltonian}
H = -\Jxy\sum_{i = 1}^3 (S_i^x S_{i+1}^x + S_i^y S_{i+1}^y) -
\Jz \sum_{i = 1}^3 S_i^z S_{i+1}^z + H_Z,
\end{equation}
where $\bsy{S}_4 = \bsy{S}_1$. Here,
$\Jxy$ and $\Jz$ are the in- and out-of-plane exchange coupling
constants, respectively, and $H_Z = \sum_{i = 1}^3
\mathbf{\b}_i \cdot \bsy{S}_i$ denotes the Zeeman coupling of the spins
$\bsy{S}_i$ to the externally applied magnetic fields $\bsy{b}_i$ at
the sites~$i$ \cite{footnote01}. We now seek a configuration of $\bsy{b}_i$'s yielding a highly
entangled GHZ- or W-type \textit{ground state}. Finite-temperature
effects will then be studied in a second step.

\emph{Ground-state properties}.--- We first consider
isotropic exchange couplings, i.e.,  $\Jxy = \Jz \equiv J$. For $\bsy{b}_i = 0$, we naturally find 
two fourfold-degenerate eigenspaces due to the high symmetry of the system.  
For $J > 0$, i.e., ferromagnetic coupling, the ground-state quadruplet is
spanned by the two GHZ states $|\mathrm{GHZ}^\pm\rangle$, the W and the spin-flipped W state. 
Appropriately chosen magnetic fields allow one, however, to split off an approximate
GHZ state from this degenerate eigenspace. To identify the optimal
field geometry, we first observe that the two states
$|\mathrm{GHZ}^\pm\rangle$ have the form of a tunnel doublet. If we
thus find a set of $\bsy{b}_i$'s, which, in the classical spin system, results in precisely two degenerate
minima for the configurations
${\uparrow\uparrow\uparrow}$ and ${\downarrow\downarrow\downarrow}$
with an energy barrier in between, quantum tunneling will yield the
desired states. In order to single out exactly the two directions
perpendicular to the $xy$-plane, the magnetic fields must be
in-plane, be of the same strength, and sum to zero. This immediately implies that successive directions of the fields must differ by an angle of $2\pi/3$ from each other. 
We choose the fields to point radially outwards, although any other configuration possessing 
the required symmetry is equivalent. However, this setup is experimentally most feasible, e.g.,
by placing a bar magnet below the center of the sample (in the case of a solid state implementation).
In order to favor parallel spin configurations we consider the regime where $J \gg b$, $b = |\bsy{b}_i|$ being the Zeeman energy.
We may thus assume that for given mean spherical angles $\bar\vartheta$ (zenith) and $\bar\varphi$ (azimuth), the orientation of each spin will deviate from these values only by a small amount. Expanding the classical energy $E_c(\bar\vartheta, \bar\varphi)$ corresponding to Eq.~\eqref{anisotropic hamiltonian} to second order in these deviations and minimizing with respect to them under the constraint that they separately sum to zero yields:
\begin{equation}\label{classical energy}
E_c \approx -\frac{(b/J)^2}{8}(3 + \cos{2\bar\vartheta}) + \frac{(b/J)^3}{24}\sin{(3\bar\varphi)}\sin^3\bar\vartheta.
\end{equation}
This expression is minimal for $\bar\vartheta = 0$ and $\bar\vartheta = \pi$, representing the desired configurations. The paths in $\bar\vartheta$ with lowest barrier height connecting these two minima are found for values of $\bar\varphi = -\pi/6 + 2\pi n/3 \mod 2\pi$, $n = 0, 1, 2$, reflecting the rotational symmetry of the system. The corresponding barrier height is approximately given by $[(b/J)^2 -(b/J)^3/6]/4$ \cite{footnote02}.

\begin{figure}[t!]
 \begin{center}
 \includegraphics[width=0.8\columnwidth]{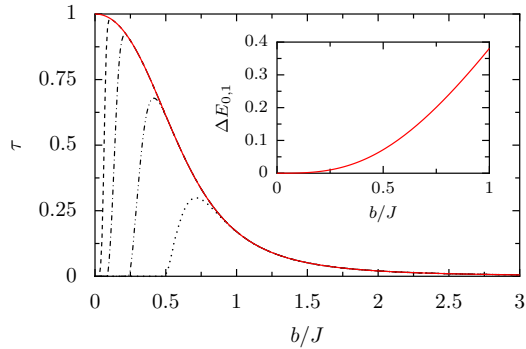}
 \caption{(color online) The tangle~$\tau$ of the system with isotropic positive
   (ferromagnetic) coupling $J$ and radial magnetic field as a
   function of $b/J$ for different temperatures $T = 10^{-4} J/\kb$
   (dashed line), $10^{-3} J/\kb$ (dash-dotted line), $10^{-2} J/\kb$
   (dash-dot-dotted line) and $5\times 10^{-2} J/\kb$ (dotted line). Note the twofold influence of the temperature on $\tau$: Although higher temperatures reduce the maximally achievable entanglement, a stabilizing effect is observed as well. A maximum in the tangle is more robust against fluctuations in $b$ at higher temperatures due to the less rapid drop-off of $\tau$ as $b/J$ is reduced. Conversely, $\tau$ of the approximate GHZ g.s.~$|\mathrm{GS}\rangle$ ($T = 0$, red line) shows a discontinuity at $b = 0$, where $\tau(\rho) = 0$. For ${b>0}$, we find the simple algebraic expression
   $\tau_p(|\mathrm{GS}\rangle) = (3 - 8b/J)/C + 2/\sqrt{C}$, where $C = 9 + 4b(4b/J - 3)/J$.
   Inset: Energy splitting $\Delta E_{0,1}$ of the ground-state
   doublet as a function of $b/J$.}
 \label{fig:tangle_isotropic}
 \end{center}
\end{figure}

Next we return to the quantum system. The Hamiltonian~\eqref{anisotropic hamiltonian} with isotropic exchange
coupling $J$ and radial magnetic field can be diagonalized exactly. Expanding for
${\b/J \ll 1}$, the overlap probabilities of
the exact ground  state $|0\rangle$ with $|\mathrm{GHZ}^{+}\rangle_{\mathrm{l.u.}}$ and the exact first excited state $|1\rangle$ with $|\mathrm{GHZ}^{-}\rangle_{\mathrm{l.u.}}$, respectively, are identical to second order and are given by ${|
{}_{\mathrm{l.u.}}\langle\mathrm{GHZ^{+}}|0\rangle|^2} = {|
{}_{\mathrm{l.u.}}\langle\mathrm{GHZ^{-}}|1\rangle|^2} \approx 1 - \frac{1}{3} (\b/J)^2$ (`l.u.' indicates that the states are equivalent to GHZ states via local unitary transformations). The associated
energy splitting is given by $\Delta
E_{0,1}\approx 2(\b/J)^3/3$ (see inset of Fig.~\ref{fig:tangle_isotropic}). This confirms the above semiclassical
considerations in terms of tunnel doublets. Moreover, we see that the g.s.~can only approximate a GHZ state although this approximation will turn out to be very good even at finite temperatures
where mixing with excited states additionally decreases the entanglement.
Before discussing this in greater detail, we study the ground-state of the
general anisotropic case with $\Jxy \neq \Jz$ in the Hamiltonian~\eqref{anisotropic hamiltonian}.

When $\Jxy \neq \Jz$ it is possible to generate highly entangled states by
applying a spatially uniform magnetic field either perpendicular to
or in the $xy$-plane. Indeed, a field along the $z$-axis, i.e.,
$\bsy{b}_i = b\bsy{e}_z$, $i = 1,2,3,$ with $\b > 0$ yields an exact W state
as g.s.~if $\Jxy > 0$ and $b < \Jxy - \Jz$ (note that this implies the condition $\Jxy > \Jz$). The
optimal Zeeman energy $b_{\mathrm{opt}}$ leading to the highest energy
splitting $\Delta E_\mathrm{opt}$ between the g.s.~and the first excited state is given by $b_\mathrm{opt} = (\Jxy - \Jz)/2$. This yields $\Delta E_\mathrm{opt} = 3\Jxy /2$ if $\Jz < -2\Jxy$ and $\Delta E_\mathrm{opt} = (\Jxy - \Jz)/2$ otherwise.  
The W state is thus best realized by choosing $b = b_\mathrm{opt}$ together with a temperature sufficiently small compared to $\Delta E_\mathrm{opt}$. In order to obtain a GHZ state,
one has to apply an in-plane magnetic field $\bsy{b}_i = b\bsy{e}_x$.
In this case we find for ${\Jz > 0}$, ${-2\Jz < \Jxy < \Jz}$ a situation
similar to the one in the case of isotropic coupling and radial
magnetic field: The g.s.~converges to a GHZ state for
vanishing field but also the energy difference to the first excited
state goes to zero in this limit.

\emph{Entanglement measure}.--- Below, we will quantitatively
study the effects of finite temperature $T > 0$ on the amount of
entanglement present in the system. For
this purpose, we evaluate a suitable mixed-state EM of the canonical density matrix $\rho$ of the system. The
three-tangle, or simply \emph{tangle} $\tau_p$ (originally called
residual entanglement), is an EM for pure states $|\psi\rangle \in
\mathcal{H}_1\otimes\mathcal{H}_2\otimes\mathcal{H}_3$ of three qubits. It reads \cite{Coffman2000}
\begin{equation}
\tau_p(|\psi\rangle) = 4\det\trace_{2,3}\rho_p -
\mathcal{C}^2(\trace_{3}\rho_p) - \mathcal{C}^2(\trace_{2}\rho_p),
\end{equation}
where $\rho_p = |\psi\rangle\langle\psi |$, $\trace_i$ denotes the partial trace over subsystem $i$, and
$\mathcal{C}$
is the two-qubit concurrence \cite{Wootters1998}. The tangle takes values between $0$ and $1$ and is maximal for
GHZ states. It is also known that $\tau_p$ is an entanglement monotone \cite{Dur2000}. The generalization of pure-state monotones to mixed states is given by the so-called convex roof \cite{Uhlmann2000, Mintert2005, Lohmayer2006}. Accordingly, the mixed-state tangle $\tau$ is defined as
\begin{equation}\label{tangle mixed}
\tau(\rho) = \inf_{\{p_i,
|\psi_i\rangle\}\in\mathfrak{D}(\rho)}\sum_i p_i
\tau_p(|\psi_i\rangle).
\end{equation}
Here, $\mathfrak{D}(\rho)$ denotes the set of all pure-state
decompositions $\{p_i,|\psi_i\rangle\}_{i = 1}^K$ of $\rho$, with $\sum_{i=1}^K p_i
|\psi_i\rangle\langle\psi_i| = \rho$, $p_i \geq 0$, $\sum_{i =
1}^K p_i = 1$ and $K \geq R = \rank \rho$. The above definition of $\tau$ ensures that $\tau(\rho) = \tau_p(|\psi\rangle)$ if $\rho = |\psi\rangle\langle\psi|$, and that $\tau$ itself is an entanglement monotone \cite{Mintert2005}.

\emph{Numerical evaluation}.--- 
In order to tackle the optimization problem in Eq. \eqref{tangle
mixed} numerically, the set of all pure-state decompositions
$\mathfrak{D}(\rho)$ needs to be given in an explicitly
parameterized form. It is known \cite{Hughston1993, Kirkpatrick2005}
that every pure state decomposition $\{p_i,|\psi_i\rangle\}_{i =
1}^K$ of $\rho$ is related to a complex $K\times R$ matrix $U$ satisfying the unitary constraint
$U^\dag U = \mathds{1}_{R\times R}$, i.e., a matrix having orthonormal column vectors \cite{footnote03}. 
In fact, the set of
all such matrices, the so-called Stiefel manifold $St(K, R)$, provides a complete parametrization of all
pure-state decompositions $\{p_i,|\psi_i\rangle\}_{i = 1}^K\in
\mathfrak{D}(\rho)$ of $\rho$ with fixed cardinality $K$.
The minimization problem in Eq. \eqref{tangle mixed} can thus be rewritten as
\begin{equation}\label{general problem}
f(\rho) = \min_{K \geq R}\inf_{U\in St(K, R)} h(U, \rho),
\end{equation}
where in our case $h$ is the sum over the weighted pure-state tangles with probabilities
and state vectors obtained from $\rho$ via the matrix $U$. Problems of this kind are considered to be extremely difficult to solve
in general \cite{Plenio2007}.
We have performed the minimization over the Stiefel manifold numerically using the method described below. We have found that
the thereby obtained values converge quickly as $K$ is increased, and have thus fixed $K = R +
4$ throughout all of our calculations, yielding an accuracy which is by far sufficient for our purpose (note
that decompositions with smaller cardinality are contained as well).
The numerical method we used is a generalization of the conjugate gradient algorithm presented in Ref. \cite{Audenaert2001}. It is however
only suited for searching over the unitary manifold $St(K, K)$. At the cost of over-parameterizing the search space, we have to minimize over $K\times K$ matrices using only the first $R$ columns. The iterative algorithm builds conjugate search directions $X$ (skew-hermitian $K\times K$ matrices) from the gradient $G$ at the current iteration point $U$ and the previous search direction using a modified Polak-Ribi\`{e}re update formula. A line search along the geodesic $g(t) = U\exp(tX)$ going through $U$ in direction $X$ is performed in every step. In Ref. \cite{Audenaert2001}, an analytical expression for the gradient $G$ is given in the case where $f$ is the EOF. The algorithm is however also applicable to a generic convex-roof EM $f$ of the form \eqref{general problem}. 
We find the matrix elements $G_{jk}$ of the general gradient $G$ to be
\begin{equation}
G_{jk} = (A_{jk} - A_{kj})/2 + \mathfrak{i}(S_{jk} + S_{kj})/2,
\end{equation}
where
\begin{align}
A_{jk} &= \sum_{i=1}^K \left(\frac{\partial h}{\partial\re U_{ik}}\re U_{ij} + \frac{\partial h}{\partial\im U_{ik}}\im U_{ij} \right), \\
S_{jk} &= \sum_{i=1}^K \left(\frac{\partial h}{\partial\im U_{ik}}\re U_{ij} - \frac{\partial h}{\partial\re U_{ik}}\im U_{ij} \right).
\end{align}
The derivatives of $h$ with respect to the real and imaginary parts of $U_{ik}$, $\re U_{ik}$ and $\im U_{ik}$, respectively, are taken at $U$ and can be evaluated numerically using finite differences. We have tested our implementation by
comparing our numerical results to known analytical results. The maximal encountered absolute error was smaller than $10^{-13}$ for the EOF of isotropic $2\times 2$ states \cite{Terhal2000}, $10^{-12}$ for $3 \times 3$ states and $10^{-10}$ for the tangle of a GHZ/W mixture \cite{Lohmayer2006}. This suggests that, although our method can only provide an upper bound, this bound is very tight. It was shown only recently that also a (typically tight) lower bound on any entanglement monotone can be estimated using entanglement witnesses \cite{Guhne2007, Eisert2007}. This is an interesting subject which is left for future research. 

\textit{Finite temperature}.--- We return to the study of the three
qubits described by the Hamiltonian \eqref{anisotropic hamiltonian}.
Using the generalized conjugate gradient algorithm, we are able to
investigate the entanglement as a function of the temperature~$T$, the
magnetic field strength $\b$ and the exchange couplings $\Jxy$ and
$\Jz$ by calculating the mixed-state tangle $\tau(\rho)$, where $\rho =
\exp(-H/\kb T)/\trace \exp(-H/\kb T)$ is the canonical density matrix of the system. To our knowledge, this is the first time that $\tau (\rho)$ has been evaluated for states arising from a physical model. Our main goal now is to maximize the entanglement as a function of $b \equiv |\bsy{b}_i|$, i.e., the Zeeman energy. For this purpose we consider only GHZ states in the following, since our W ground states are $b$-independent (see above).
In the system with isotropic
exchange coupling $J > 0$ and radial magnetic field, the tangle $\tau$ tends
to zero for ${b/J \rightarrow 0}$ due to the vanishing energy
splitting $\Delta E_{0, 1}$ (see
Fig.~\ref{fig:tangle_isotropic}). We remark that this behavior is discontinuous at $T = 0$, where $\tau(\rho) \rightarrow 1$ for $b/J \rightarrow 0$, but $\tau(\rho) = 0$ at $b = 0$. With larger
$b/J$, the g.s.~contributes dominantly to $\rho$ but simultaneously deviates increasingly from a GHZ state. The entanglement in the system is therefore reduced (cf.\ solid line in
Fig.~\ref{fig:tangle_isotropic}).  For a given temperature, the
maximal tangle~$\tau_\mathrm{max}$ is therefore obtained at a finite
optimal value~$(\b/J)_\mathrm{opt}$ of the scaled magnetic field strength as a
trade-off between having a highly entangled g.s.~and separating
the latter from excited states in order to avoid the negative
effects of mixing. For low temperatures $T \lesssim 10^{-2} J/\kb$, we
numerically find the power laws $(\b/J)_\mathrm{opt} \propto (\kb T/
J)^{\alpha}$ and $1 - \tau_\mathrm{max} \propto (\kb T/J)^\beta$ with
the exponents $\alpha\approx0.30$ and $\beta\approx0.63$.  
Specifically, we obtain $\tau(\rho) = 0.98$~$(0.92)$ for $T = 10^{-4} J/\kb$~$(10^{-3} J/\kb)$ and $b = 0.11 J$~$(0.21 J)$. Apart from the effect of reducing $\tau_\mathrm{max}$, finite temperatures also possess the advantageous feature of broadening the discontinuity of $\tau$ at $T = 0$ and $b = 0$ which makes $\tau_\mathrm{max}$ more stable against fluctuations of $b$ around $\b_\mathrm{opt}$ (see Fig.~\ref{fig:tangle_isotropic}). 

\begin{figure}[t]
  \centering
 \includegraphics[width=0.8\columnwidth]{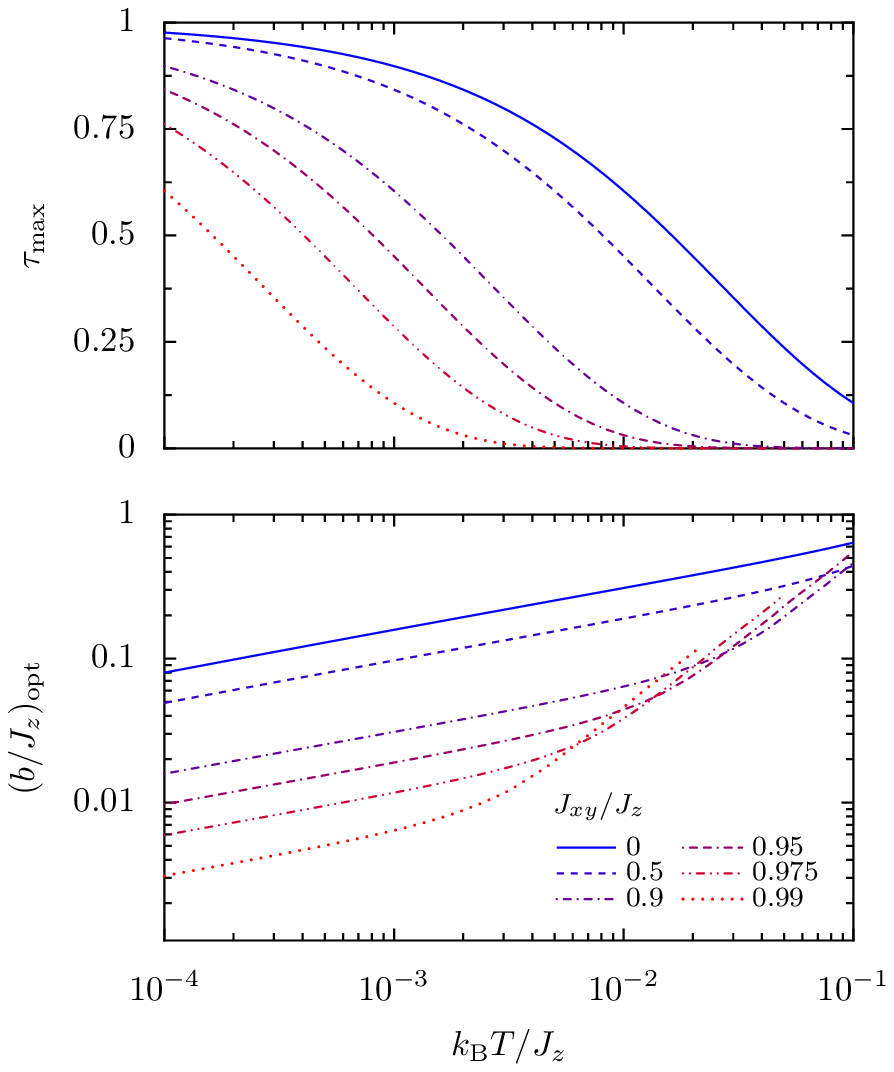}
 \caption{(color online) Top: Maximally achievable tangle
   $\tau_\mathrm{max}$ in the anisotropic system (GHZ g.s.) with homogeneous
   {in-plane} magnetic field and $\Jz > 0$ as a function of
   temperature for six anisotropy ratios $J_{xy}/J_z$ (see
   legend). The curves end at $\tau(\rho) = 10^{-5}$. Bottom: The corresponding optimal values
   $(\b/J_z)_\mathrm{opt}$ of the scaled magnetic field strength~$b/\Jz$.}
 \label{fig:tangle_anisotropic}
\end{figure}

We finally come back to the general anisotropic model~\eqref{anisotropic hamiltonian} 
with $\Jxy \neq \Jz$ subject to a homogeneous in-plane magnetic field.  In
Fig.~\ref{fig:tangle_anisotropic} we show the maximally achievable
tangle~$\tau_\mathrm{max}$ (optimized with respect to $\b/\Jz$) as a function of temperature for
various anisotropy ratios $\Jxy/\Jz$ (where, as before, $\Jz > 0$). Since we are interested in high values of $\tau_\mathrm{max}$, an arbitrary but low cutoff was introduced in the calculation at $\tau(\rho) = 10^{-5}$. The lower panel of Fig.~\ref{fig:tangle_anisotropic} depicts the corresponding optimal field values~$(\b/\Jz)_\mathrm{opt}$. At low temperatures $T$, a power-law dependence of $(\b/\Jz)_\mathrm{opt}$ on
$T$ is observed, similar to the above isotropic case.
Note that a higher amount of entanglement can be realized in
systems with stronger anisotropies. E.g., for
Ising coupling ($\Jxy/\Jz=0$) we find $\tau(\rho) = 0.98$~$(0.89)$ for $T = 10^{-4} J/\kb$~$(10^{-3} J/\kb)$ and $b = 0.080 \Jz$~$(0.16 \Jz)$. At $T = 10^{-4} J/\kb$ but with $\Jxy/\Jz = 0.9$, still a very good value $\tau(\rho) = 0.90$ is achieved for $b = 0.016 \Jz$. We remark that still higher tangles are obtained for negative (antiferromagnetic) $\Jxy > -2\Jz$. In this case, the maximal tangle as a function of $T$ decays even more slowly than the curves displayed in the top panel of Fig.~\ref{fig:tangle_anisotropic}.

Possible implementations of the qubits include GaAs and InAs quantum dots, InAs nanowires or single-wall carbon nanotubes. Assuming a typical value of $|J| \sim 1\,\mathrm{meV}$ \cite{Cerletti2005, Hanson2007} we obtain $\tau \approx 0.9$ at $T \approx 10\,\mathrm{mK}$ and $B \approx 2\,\mathrm{T}$ (assuming a $g$-factor of $|g| = 2$). Ferromagnetic coupling is achieved by operating the dots with more than one electron per dot.


We thank D. Bulaev, G. Burkard, L. Chirolli, W.A. Coish, and D.
Stepanenko for useful discussions. Financial support by the EU RTN
QuEMolNa, the EU NoE MAGMANet, the NCCR Nanoscience, and the Swiss NSF
is acknowledged.

\bibliographystyle{apsrev}

\begin{thebibliography}{19}
\expandafter\ifx\csname natexlab\endcsname\relax\def\natexlab#1{#1}\fi
\expandafter\ifx\csname bibnamefont\endcsname\relax
  \def\bibnamefont#1{#1}\fi
\expandafter\ifx\csname bibfnamefont\endcsname\relax
  \def\bibfnamefont#1{#1}\fi
\expandafter\ifx\csname citenamefont\endcsname\relax
  \def\citenamefont#1{#1}\fi
\expandafter\ifx\csname url\endcsname\relax
  \def\url#1{\texttt{#1}}\fi
\expandafter\ifx\csname urlprefix\endcsname\relax\def\urlprefix{URL }\fi
\providecommand{\bibinfo}[2]{#2}
\providecommand{\eprint}[2][]{\url{#2}}

\bibitem{Coll01}
\bibinfo{author}{\bibfnamefont{A.}~\bibnamefont{Einstein}},
  \bibinfo{author}{\bibfnamefont{B.}~\bibnamefont{Podolsky}}, \bibnamefont{and}
  \bibinfo{author}{\bibfnamefont{N.}~\bibnamefont{Rosen}},
  \bibinfo{journal}{Phys. Rev.} \textbf{\bibinfo{volume}{47}},
  \bibinfo{pages}{777} (\bibinfo{year}{1935}); \bibinfo{author}{\bibfnamefont{J.}~\bibnamefont{Bell}},
  \bibinfo{journal}{Physics} \textbf{\bibinfo{volume}{1}}, \bibinfo{pages}{195}
  (\bibinfo{year}{1964}); \bibinfo{author}{\bibfnamefont{A.}~\bibnamefont{Aspect}},
  \bibinfo{author}{\bibfnamefont{P.}~\bibnamefont{Grangier}}, \bibnamefont{and}
  \bibinfo{author}{\bibfnamefont{G.}~\bibnamefont{Roger}},
  \bibinfo{journal}{Phys. Rev. Lett.} \textbf{\bibinfo{volume}{49}},
  \bibinfo{pages}{91} (\bibinfo{year}{1982}); \bibinfo{author}{\bibfnamefont{R.~F.} \bibnamefont{Werner}},
  \bibinfo{journal}{Phys. Rev. A} \textbf{\bibinfo{volume}{40}},
  \bibinfo{pages}{4277} (\bibinfo{year}{1989}).

\bibitem{Greenberger1989}
\bibinfo{author}{\bibfnamefont{D.~M.} \bibnamefont{Greenberger}},
  \bibinfo{author}{\bibfnamefont{M.}~\bibnamefont{Horne}}, \bibnamefont{and}
  \bibinfo{author}{\bibfnamefont{A.}~\bibnamefont{Zeilinger}},
  \emph{\bibinfo{title}{Bell's Theorem, Quantum Theory, and Conceptions of the
  Universe}} (\bibinfo{publisher}{Kluwer Academic Publishers, Dortrecht},
  \bibinfo{year}{1989}).

\bibitem[{\citenamefont{D\"ur et~al.}(2000)\citenamefont{D\"ur, Vidal, and
  Cirac}}]{Dur2000}
\bibinfo{author}{\bibfnamefont{W.}~\bibnamefont{D\"ur}},
  \bibinfo{author}{\bibfnamefont{G.}~\bibnamefont{Vidal}}, \bibnamefont{and}
  \bibinfo{author}{\bibfnamefont{J.~I.} \bibnamefont{Cirac}},
  \bibinfo{journal}{Phys. Rev. A} \textbf{\bibinfo{volume}{62}},
  \bibinfo{pages}{062314} (\bibinfo{year}{2000}).

\bibitem[{\citenamefont{Nielsen and Chuang}(2000)}]{Nielsen2000a}
\bibinfo{author}{\bibfnamefont{M.~A.} \bibnamefont{Nielsen}} \bibnamefont{and}
  \bibinfo{author}{\bibfnamefont{I.~L.} \bibnamefont{Chuang}},
  \emph{\bibinfo{title}{Quantum Computation and Quantum Information}}
  (\bibinfo{publisher}{Cambridge University Press}, \bibinfo{address}{New
  York}, \bibinfo{year}{2000}).

\bibitem[{\citenamefont{Raussendorf et~al.}(2003)\citenamefont{Raussendorf,
  Browne, and Briegel}}]{Raussendorf2003}
\bibinfo{author}{\bibfnamefont{R.}~\bibnamefont{Raussendorf}},
  \bibinfo{author}{\bibfnamefont{D.~E.} \bibnamefont{Browne}},
  \bibnamefont{and} \bibinfo{author}{\bibfnamefont{H.~J.}
  \bibnamefont{Briegel}}, \bibinfo{journal}{Phys. Rev. A}
  \textbf{\bibinfo{volume}{68}}, \bibinfo{pages}{022312}
  (\bibinfo{year}{2003}).


\bibitem[{\citenamefont{Bose}(2003)}]{Bose2003}
\bibinfo{author}{\bibfnamefont{S.}~\bibnamefont{Bose}}, \bibinfo{journal}{Phys.
  Rev. Lett.} \textbf{\bibinfo{volume}{91}}, \bibinfo{pages}{207901}
  (\bibinfo{year}{2003}).


\bibitem[{\citenamefont{Rajagopal and Rendell}(2002)}]{Rajagopal2002}
\bibinfo{author}{\bibfnamefont{A.~K.} \bibnamefont{Rajagopal}}
  \bibnamefont{and} \bibinfo{author}{\bibfnamefont{R.~W.}
  \bibnamefont{Rendell}}, \bibinfo{journal}{Phys. Rev. A}
  \textbf{\bibinfo{volume}{65}}, \bibinfo{pages}{032328}
  (\bibinfo{year}{2002}).

\bibitem[{\citenamefont{Plenio and Virmani}(2007)}]{Plenio2007}
\bibinfo{author}{\bibfnamefont{M.~B.} \bibnamefont{Plenio}} \bibnamefont{and}
  \bibinfo{author}{\bibfnamefont{S.}~\bibnamefont{Virmani}},
  \bibinfo{journal}{Quant. Inf. Comp.} \textbf{\bibinfo{volume}{7}},
  \bibinfo{pages}{1} (\bibinfo{year}{2007}).

\bibitem{Mintert2005} F. Mintert {\it et al.}, Phys. Rep. \textbf{415}, 207 (2005).

\bibitem{Bennett1996} C.~H. Bennett {\it et al.}, Phys. Rev. A \textbf{54}, 3824 (1996).

\bibitem[{\citenamefont{Coffman et~al.}(2000)\citenamefont{Coffman, Kundu, and
  Wootters}}]{Coffman2000}
\bibinfo{author}{\bibfnamefont{V.}~\bibnamefont{Coffman}},
  \bibinfo{author}{\bibfnamefont{J.}~\bibnamefont{Kundu}}, \bibnamefont{and}
  \bibinfo{author}{\bibfnamefont{W.~K.} \bibnamefont{Wootters}},
  \bibinfo{journal}{Phys. Rev. A} \textbf{\bibinfo{volume}{61}},
  \bibinfo{pages}{052306} (\bibinfo{year}{2000}).

\bibitem[{\citenamefont{Wei and Goldbart}(2003)}]{Wei2003}
\bibinfo{author}{\bibfnamefont{T.-C.} \bibnamefont{Wei}} \bibnamefont{and}
  \bibinfo{author}{\bibfnamefont{P.~M.} \bibnamefont{Goldbart}},
  \bibinfo{journal}{Phys. Rev. A} \textbf{\bibinfo{volume}{68}},
  \bibinfo{pages}{042307} (\bibinfo{year}{2003}).

\bibitem[{\citenamefont{Audenaert et~al.}(2001)\citenamefont{Audenaert,
  Verstraete, and De~Moor}}]{Audenaert2001}
\bibinfo{author}{\bibfnamefont{K.}~\bibnamefont{Audenaert}},
  \bibinfo{author}{\bibfnamefont{F.}~\bibnamefont{Verstraete}},
  \bibnamefont{and} \bibinfo{author}{\bibfnamefont{B.}~\bibnamefont{De~Moor}},
  \bibinfo{journal}{Phys. Rev. A} \textbf{\bibinfo{volume}{64}},
  \bibinfo{pages}{052304} (\bibinfo{year}{2001}).

\bibitem[{\citenamefont{Uhlmann}(2000)}]{Uhlmann2000}
\bibinfo{author}{\bibfnamefont{A.}~\bibnamefont{Uhlmann}},
  \bibinfo{journal}{Phys. Rev. A} \textbf{\bibinfo{volume}{62}},
  \bibinfo{pages}{032307} (\bibinfo{year}{2000}).

\bibitem{footnote01} Depending on the actual implementation of the qubits, $\bsy{b}_i$ can denote an effective magnetic field.

\bibitem{footnote02} Using semiclassical path integration techniques \cite{Loss1992}, we can calculate the tunnel splitting from Eq. \eqref{classical energy}. However, such a procedure gives accurate results only for large spins ($S\gg 1$) and is thus not pursued here.

\bibitem[{\citenamefont{Loss et~al.}(1992)\citenamefont{Loss, DiVincenzo, and
  Grinstein}}]{Loss1992}
\bibinfo{author}{\bibfnamefont{D.}~\bibnamefont{Loss}},
  \bibinfo{author}{\bibfnamefont{D.~P.} \bibnamefont{DiVincenzo}},
  \bibnamefont{and}
  \bibinfo{author}{\bibfnamefont{G.}~\bibnamefont{Grinstein}},
  \bibinfo{journal}{Phys. Rev. Lett.} \textbf{\bibinfo{volume}{69}},
  \bibinfo{pages}{3232} (\bibinfo{year}{1992}).

\bibitem[{\citenamefont{Wootters}(1998)}]{Wootters1998}
\bibinfo{author}{\bibfnamefont{W.~K.} \bibnamefont{Wootters}},
  \bibinfo{journal}{Phys. Rev. Lett.} \textbf{\bibinfo{volume}{80}},
  \bibinfo{pages}{2245} (\bibinfo{year}{1998}).

\bibitem{Lohmayer2006} R. Lohmayer {\it et al.}, Phys. Rev. Lett. \textbf{97}, 260502 (2006).

\bibitem[{\citenamefont{Hughston et~al.}(1993)\citenamefont{Hughston, Jozsa,
  and Wootters}}]{Hughston1993}
\bibinfo{author}{\bibfnamefont{L.~P.} \bibnamefont{Hughston}},
  \bibinfo{author}{\bibfnamefont{R.}~\bibnamefont{Jozsa}}, \bibnamefont{and}
  \bibinfo{author}{\bibfnamefont{W.~K.} \bibnamefont{Wootters}},
  \bibinfo{journal}{Phys. Lett. A} \textbf{\bibinfo{volume}{183}},
  \bibinfo{pages}{14} (\bibinfo{year}{1993}).

\bibitem[{\citenamefont{Kirkpatrick}(2005)}]{Kirkpatrick2005}
\bibinfo{author}{\bibfnamefont{K.~A.} \bibnamefont{Kirkpatrick}},
  \bibinfo{journal}{Found. Phys. Lett.} \textbf{\bibinfo{volume}{19}},
  \bibinfo{pages}{95} (\bibinfo{year}{2005}).
  
\bibitem{footnote03} Given $\rho$ and $U$ with $U^\dag U = \mathds{1}_{R\times R}$, $\{p_i,|\psi_i\rangle\}_{i =
1}^K$ is obtained as $p_i = \langle\tilde\psi_i|\tilde\psi_i\rangle$, $|\psi_i\rangle = (1/\sqrt{p_i})|\tilde\psi_i\rangle$, where $|\tilde\psi_i\rangle = \sum_{j=1}^R U_{ij}\sqrt{\lambda_j}|\chi_j\rangle$ and $|\chi_i\rangle$ are the $R$ eigenvectors of $\rho$ with non-zero eigenvalues $\lambda_i$.

\bibitem[{\citenamefont{Terhal and Vollbrecht}(2000)}]{Terhal2000}
\bibinfo{author}{\bibfnamefont{B.~M.} \bibnamefont{Terhal}} \bibnamefont{and}
  \bibinfo{author}{\bibfnamefont{K.~G.~H.} \bibnamefont{Vollbrecht}},
  \bibinfo{journal}{Phys. Rev. Lett.} \textbf{\bibinfo{volume}{85}},
  \bibinfo{pages}{2625} (\bibinfo{year}{2000}).

\bibitem[{\citenamefont{G\"uhne et~al.}(2007)\citenamefont{G\"uhne, Reimpell,
  and Werner}}]{Guhne2007}
\bibinfo{author}{\bibfnamefont{O.}~\bibnamefont{G\"uhne}},
  \bibinfo{author}{\bibfnamefont{M.}~\bibnamefont{Reimpell}}, \bibnamefont{and}
  \bibinfo{author}{\bibfnamefont{R.~F.} \bibnamefont{Werner}},
  \bibinfo{journal}{Phys. Rev. Lett.} \textbf{\bibinfo{volume}{98}},
  \bibinfo{eid}{110502} (\bibinfo{year}{2007}).

\bibitem[{\citenamefont{Eisert et~al.}(2007)\citenamefont{Eisert, Brand\~{a}o,
  and Audenaert}}]{Eisert2007}
\bibinfo{author}{\bibfnamefont{J.}~\bibnamefont{Eisert}},
  \bibinfo{author}{\bibfnamefont{F.~G. S.~L.} \bibnamefont{Brand\~{a}o}},
  \bibnamefont{and} \bibinfo{author}{\bibfnamefont{K.~M.~R.}
  \bibnamefont{Audenaert}}, \bibinfo{journal}{New J. Phys.}
  \textbf{\bibinfo{volume}{9}}, \bibinfo{pages}{46} (\bibinfo{year}{2007}).

\bibitem{Cerletti2005} V. Cerletti {\it et al.}, {Nano\-technology} \textbf{16}, R27 (2005).
  
\bibitem{Hanson2007} R. Hanson {\it et al.}, Rev. Mod. Phys. \textbf{79}, 1217 (2007).



\end{thebibliography}

\end{document}